\documentclass[11pt]{article}
\usepackage{amsmath,amssymb,color,graphics,epsfig,cite}

\usepackage{amsfonts}

\newcommand{\hoch}[1]{$\, ^{#1}$}

\makeatletter
\@addtoreset{equation}{section}
\makeatother

\newcommand{\be}{\begin{equation}}
\newcommand{\ee}{\end{equation}}
\newcommand{\bea}{\setlength\arraycolsep{2pt} \begin{eqnarray}}
\newcommand{\eea}{\end{eqnarray}}
\newcommand{\nn}{\nonumber}

\def\ft#1#2{{\textstyle{\frac{\scriptstyle #1}{\scriptstyle #2} } }}
\def\fft#1#2{{\frac{#1}{#2}}}

\def\0{{\sst{(0)}}}
\def\1{{\sst{(1)}}}
\def\2{{\sst{(2)}}}
\def\3{{\sst{(3)}}}
\def\4{{\sst{(4)}}}
\def\5{{\sst{(5)}}}
\def\6{{\sst{(6)}}}
\def\7{{\sst{(7)}}}
\def\8{{\sst{(8)}}}
\def\sst#1{{\scriptscriptstyle #1}}

\begin{document}


\begin{center}
{\Large {\bf Einstein-Vector Gravity, Emerging Gauge Symmetry \\ and de Sitter Bounce}}

\vspace{40pt}
{\bf Wei-Jian Geng and H. L\"u}

\vspace{10pt}

\hoch{1}{\it Center for Advanced Quantum Studies, Department of Physics, \\
Beijing Normal University, Beijing 100875, China}

\vspace{40pt}

\underline{ABSTRACT}
\end{center}

We construct a class of Einstein-vector theories where the vector field couples bilinearly to the curvature polynomials of arbitrary order in such a way that only Riemann tensor rather than its derivative enters the equations of motion.  The theories can thus be ghost free. The $U(1)$ gauge symmetry may emerge in the vacuum and also in some weak-field limit. We focus on the two-derivative theory and study a variety of applications.  We find that in this theory, the energy-momentum tensor of dark matter provides a position-dependent gauge-violating term to the Maxwell field. We also use the vector as an inflaton and construct cosmological solutions.  We find that the expansion can accelerate without a bared cosmological constant, indicating a new candidate for dark energy.  Furthermore we obtain exact solutions of de Sitter bounce, generated by the vector which behaves like a Maxwell field in the later time.  We also obtain a few new exact black holes that are asymptotic to flat and Lifshitz spacetimes.  In addition, we construct exact wormholes, and Randall-Sundrum II domain walls.

\vfill {\footnotesize gengwj@mail.bnu.edu.cn \ \ \ mrhonglu@gmail.com}

\thispagestyle{empty}

\pagebreak

\tableofcontents
\addtocontents{toc}{\protect\setcounter{tocdepth}{2}}



\section{Introduction}

Symmetries play important roles in modern physics.  They become guiding principles in constructing new theories in many branches of physics, fundamental ones in particular.  A more reductionist viewpoint typically assumes bigger symmetries in higher energy or in the ultraviolet (UV) region.  In the lower energy or in the infrared (IR) region, the symmetries may be broken via various mechanisms.  The most famous example is perhaps the spontaneous symmetry breaking by the Higgs mechanism \cite{Englert:1964et,Higgs:1964pj}.

An alternative phenomenon is that a symmetry does not in fact exist at the fundamental level, but it can emerge in the IR or weak-field limit. Many such examples of discrete global symmetries can be found in elementary particles.  These symmetries are approximate and accidental, and can be very useful for organizing the spectra.  A more non-trivial example is provided by the proposal that (continuous) Lorentz invariance might be an emergent symmetry \cite{Chadha:1982qq}.  Concrete such an example is provided by a Lifshitz scalar for some condensed matter system in the critical point. The equation has anisotropic scaling between the spatial and temporal directions.  However, under the influence of certain relevant deformation to the action, the theory flows in the infrared to recover the Lorentz invariance. (See, for example, \cite{Ardonne:2003wa,Horava:2008ih}.)

Interestingly, the symmetry breaking from the UV to IR typically deals with local ones whilst the emergent symmetries in the IR are typically global. Recently Einstein's principle of general coordinate invariance as an emergent phenomenon in the IR region was proposed in Horava-Lifshitz gravity \cite{Horava:2009uw}, although the flow to recover Einstein gravity remains to be established.

In this paper, we propose that the $U(1)$ gauge symmetry of Einstein-Maxwell gravity may be an emergent property from a more general class of Einstein-vector theories where the bared vector potential field $A=A_\mu dx^\mu$ couples to curvature tensors directly.  In other words, the general theory has no $U(1)$ gauge symmetry; however, it emerges in some vacua or some weak-field limit. An immediate objection to this proposal may be that the $U(1)$ gauge symmetry of the Maxwell theory is extremely robust and the attempts of breaking this symmetry can be easily invalidated  by experimental or observational data.  However, gravity is extremely weak compared to other fundamental forces and almost always ignored in particle physics.  Thus such a gauge-symmetry violation has no effect on elementary particle physics.  As for the behavior of the propagating light, we consider a concrete example. In addition to the Einstein-Hilbert term and the kinetic term of $A$, namely
\be
{\cal L}=\sqrt{-g} (R - \ft14 F^2)\,,\qquad F=dA\,,
\ee
we add a non-minimal coupling between gravity and the vector
\be
\sqrt{-g}\,\gamma\,G_{\mu\nu} A^\mu A^\nu\,,\label{introgaa}
\ee
where $G_{\mu\nu}=R_{\mu\nu}-\ft12 g_{\mu\nu}R$ is the Einstein tensor and $\gamma$ is some coupling constant.  In the Minkowski vacuum or some more general backgrounds such as Schwarzschild or Kerr black holes, $G_{\mu\nu}$ vanishes. The linear fluctuations of the theory in these backgrounds consist only the massless graviton and photon, and hence the $U(1)$ gauge symmetry emerges.  In our current Universe, the contribution to the spacetime curvature due to electric-magnetic fields are negligible, and hence it can be viewed as a background with vanishing $A$.  The matter energy-momentum tensor in the Einstein equation
\be
G_{\mu\nu}=T^{\rm mat}_{\mu\nu}
\ee
has mainly three sources.  The first type is the ordinary matter composed of baryons and leptons. These are typically localized, and even for distributed dust matter, the direct interactions with photons far suppress the effects through the gauge-violating term (\ref{introgaa}).  Thus our new model provides no detectable difference in the behavior of light due to these matters.  The second type is dark energy, which contributes a global (Lorentz preserving) mass term via (\ref{introgaa}) to the photon at the order of the cosmological constant.  This provides a new dispersion relation for photons and a modification to the Couloum's law. However, since the cosmological constant is extremely small, the electric static force remains effectively long-ranged and the effect on a propagating photon is detectable only if the light travels across the whole universe. It should be recalled that the van Dam-Veltman-Zakharov discontinuity of graviton \cite{vanDam:1970vg,Zakharov:1970cc} does not apply to the photon. A photon with small enough mass is indistinguishable observationally from a massless photon.
The third type is dark matter. It provides a position-dependent (and hence Lorentz violating) gauge-symmetry breaking term to the Maxwell field.  Since light does not interact dark matter directly, their interaction via the spacetime curvature
\be
\sqrt{-g}\,\gamma\,G_{\mu\nu} A^\mu A^\nu \rightarrow \sqrt{-g}\,\gamma\, T^{\rm DM}_{\mu\nu} A^\mu A^\nu\,,
\ee
may become detectable in astronomical observations. Note that this effect is over and above that light bends over the curved spacetime in its geodesic motion, for which the gauge symmetry is preserved.  Since the magnitude of this interaction is proportional to the density of the dark matter, its effect is expected to be small and may only manifest for light travels across over galactic distances, rather than within the solar system.

Background-dependent Lorentz violation can also arise in string theory. The changes of Standard Model particle velocities $\delta v$ in Type IIB string theory with D3/D7 branes are proportional to the particle energies, the D3-brane number density, and are inversely proportional to the string scale. Since the D3-brane number density can be space
dependent, one obtains the background dependent Lorentz violation \cite{Li:2011zm}.  The principle of this string motivated idea and our proposal is analogous, whilst ours have the advantage of recognising dark matter as the gauge-violating culprit and hence it is possible to make a direct contact with observational data.

We now give the outline of the paper. In section 2, we construct a general class of Einstein-vector theories in which the vector field couples bilinearly to the polynomials of curvature tensors in arbitrary order.  We find some specific combinations of the curvature polynomials analogous to Euler integrands in Lovelock gravities \cite{Lovelock:1971yv} such that only the curvature tensors rather than their derivatives appear in the full set of equations of motion.  This implies that although the theory involves more than two derivatives owing to nonlinearity, the linearized equations of motion over any background contain at most two derivatives.  Thus, for some appropriate coupling constants and in the special backgrounds, the linear perturbation can be ghost free.

The focus of the paper is on the theory where the curvature tensors enter the Lagrangian at the linear order, in which case, the theory involves at most two derivatives.  We present this theory in section 3.  We obtain the general equations of motion and discuss in some details how the $U(1)$ gauge symmetry emerges.  Although any emergent symmetry is approximate, we have the advantage of seeing how the symmetry is violated in the controlled fashion dictated by the theory rather than by hand. We discuss the possible gauge-symmetry violating terms.   In section 4, we construct some exact special black holes that are asymptotic to the Minkowski spacetime.  We discuss the properties of general black holes, of which we do not have exact solutions.  In section 5, we adopt a superpotential method to construct static solutions, and find classes of wormholes that connect two Minkowski or (anti)-de Sitter spacetimes ((A)dS).  In section 6, we construct Lifshitz spacetimes and Lifshitz black holes.  We find that the viscosity/entropy ratio is $1/(4\pi)$ for the dual field theory of these black holes.

In the usual Einstein-Maxwell theory, the vector field typically plays no role in the cosmological evolution since it would violate the cosmological principle of homogeneity and isotropy.  However, in our theory, owing to the violation of the $U(1)$ gauge symmetry, there is an extra scalar mode within the vector field.  Thus we can use the vector as an inflaton to construct Friedman-Lema\^itre-Robertson-Walker (FLRW) cosmological solutions.  Interestingly in some of our cosmological solutions, the acceleration of the expansion of a universe can occur without a bared cosmological constant. Thus the theory provides a new candidate for the dark energy.  To construct a realistic inflation model, we would like to require that after the inflation, the vector as an inflaton vanish, and that consequently the $U(1)$ gauge symmetry emerge at the linear perturbative level.  We find in section 7 that such solutions indeed exist.  This proposal has the advantage of being economical with the field content of a theory and also addresses the issue where the inflaton comes and/or goes. Furthermore we construct classes of solutions that describe de Sitter bounce.  In section 8, we obtain domain-walls that describe the Randall-Sundrum II scenario.  The paper is concluded in section 9.  Most of our solutions constructed and discussed in the main body of the paper are in four or five dimensions.  We present these solutions in general dimensions in the appendix.

\section{General Einstein-vector gravity}

In this section, we construct the general Einstein-vector theories in general $D$ dimensions.  The theories all involve only two fields, the metric $g_{\mu\nu}$ and a vector $A=A_\mu dx^\mu$.  For the vector field, we would like to require that it be linear in its equation of motion and involve at most two derivatives. This requirement is mainly for simplicity rather than some deep physical considerations, except for the fact that the Maxwell equations are linear. The gravity sector, on the other hand, is allowed to have higher-order Riemann curvature polynomials, since Einstein gravity is already a highly nonlinear theory. Such a gravity theory typically contains ghost excitations; however, there are some specific combinations called Euler integrands that yield no ghost modes. The corresponding theories are called Lovelock gravities \cite{Lovelock:1971yv}. The Euler integrands are given by
\be
E^{(k)} = \ft{1}{2^k}\, \delta_{\alpha_1\cdots\alpha_{2k}}^{\beta_1\cdots\beta_{2k}}\,
R^{\alpha_1\alpha_2}{}_{\beta_1\beta_2}\cdots R^{\alpha_{2k-1}\alpha_{2k}}{}_{\beta_{2k-1}
\beta_{2k}}\,,
\ee
where $(k)$ denotes the orders of the Riemann-tensor polynomials and
\be
\delta_{\alpha_1\cdots\alpha_{s}}^{\beta_1\cdots\beta_{s}}=
s! \delta_{[\alpha_1}^{\beta_1} \cdots \delta_{\alpha_s]}^{\beta_s}\,.
\ee
The low-lying examples are
\be
E^{(0)} =1\,,\qquad
E^{(1)}=R\,,\qquad E^{(2)} = R^2 - 4R^{\mu\nu}R_{\mu\nu} + R^{\mu\nu\rho\sigma}R_{\mu\nu\rho\sigma}\,,\quad etc.
\ee
The term $\sqrt{-g} E^{(k)}$ in the Lagrangian contributes
\be
E_{\mu}^{(k)\,\nu} = -\ft{1}{2^{k+1}} \delta_{\alpha_1\cdots\alpha_{2k}\,\mu}^{\beta_1\cdots\beta_{2k}\,\nu}\,
R^{\alpha_1\alpha_2}{}_{\beta_1\beta_2}\cdots R^{\alpha_{2k-1}\alpha_{2k}}{}_{\beta_{2k-1}
\beta_{2k}}
\ee
to the Einstein's equation of motion.  In other words, the variation
\be
\delta R^{\mu}{}_{\nu\rho\sigma} = \nabla_\rho \Gamma^{\mu}_{\sigma \nu} -
 \nabla_\sigma \Gamma^{\mu}_{\rho \nu}
\ee
yields a total derivative term in the Lagrangian and hence no Riemann-tensor factor acquires any derivative in the equations of motion.  The consequence at the linearized level is that the equations of motion involve at most two derivatives and hence the theory contains no inevitable linear ghost modes that are associated with the linear higher-derivative terms.

To couple the general higher-order curvature terms to a bared vector $A$, we find two possible quadratic structures of $A$ that have the same property as that of the Lovelock theory.  These are
\be
G^{(k)} = E^{(k)}_{\mu\nu} A^\mu A^\nu\,,\qquad
\widetilde G^{(k)} =E^{(k)} A^2\,.
\ee
Here are some low-lying examples of the $G^{(k)}$ series
\bea
G^{(0)} &=& -\ft12 A^2\,,\cr
G^{(1)} &=& G_{\mu\nu} A^\mu A^\nu\,,\qquad G_{\mu\nu}=R_{\mu\nu}-\ft12 g_{\mu\nu} R\cr
G^{(2)} &=& 2\Big(R R_{\mu\nu} - 2 R_{\mu\alpha\nu\beta} R^{\alpha\beta} +
R_{\mu\alpha\beta\gamma} R_\nu{}^{\alpha\beta\gamma} - 2 R_{\mu\alpha} R_{\nu}{}^\alpha
\cr
&&\qquad-\ft14 g_{\mu\nu} (R^{\alpha\beta\gamma\delta} R_{\alpha\beta\gamma\delta} - 4
R^{\alpha\beta}R_{\alpha\beta} + R^2)\Big)A^\mu A^\nu\,,\\
&\vdots&\nn
\eea
The Lagrangian of the full general theory is given by
\be
{\cal L}=\sqrt{-g} \Big(-\ft14 F^2 + \sum_{k=0} \big(\alpha^{(k)}\, E^{(k)} + \beta^{(k)} \widetilde G^{(k)} +
\gamma^{(k)} G^{(k)}\big)\Big)\,,\label{totalgenlag}
\ee
where $F=dA$ is the field strength of the vector potential $A$, and $F^2=F^{\mu\nu}F_{\mu\nu}$. Setting $A=0$ gives rise to pure Lovelock gravity.

We now present the field equations of motion.  The vector equation is given by
\be
\nabla_\mu F^{\mu\nu} =2\sum_{k} \big(E^{(k)} A^\nu + A^\mu E^{(k)\,\nu}_\mu\big)\,.
\ee
The Einstein equations of motion are complicated, given by
\be
\sum_{k} \Big(\alpha^{(k)}\, E_{\mu\nu}^{(k)} +
\beta^{(k)}\, \widetilde G_{\mu\nu}^{(k)} +
\gamma^{(k)}\, G_{\mu\nu}^{(k)}\Big)=\ft12 (F_{\mu\nu}^2 - \ft14 g_{\mu\nu} F^2)\,.
\ee
where $F^2_{\mu\nu}=F_{\mu\rho} F_{\nu}{}^\rho$, and
\bea
\widetilde G^{(k)}_{\mu\nu} &=& E^{(k)}_{\mu\nu} A^2 + E^{(k)} A_\mu A_\nu\cr
&&+\ft{k}{2^{k-1}} g^{\phantom{XX}}_{\beta_1(\mu}\delta_{\nu)\alpha_2\cdots\alpha_{2k}}^{\beta_1\beta_2 \cdots\beta_{2k}} R^{\alpha_3\alpha_4}{}_{\beta_3\beta_4}\cdots R^{\alpha_{2k-1}\alpha_{2k}}{}_{\beta_{2k-1}\beta_{2k}}
\nabla^{\alpha_2}\nabla_{\beta_2}A^2\,,\cr
G^{(k)}_{\mu\nu} &=& -\ft12 g_{\mu\nu} G^{(k)} + A_\rho A^{\phantom{X}}_{(\mu}
E^{(k)\rho}_{\nu)}\cr
&&+\ft{k}{2^{k+1}} \delta^{\beta_1\,\beta_2\cdots\beta_{2k}\,\sigma}_{(\mu |\alpha_2\cdots\alpha_{2k}\,\rho|} R^{\alpha_2}{}_{\nu)\beta_1\beta_1} R^{\alpha_3\alpha_4}
{}_{\beta_3\beta_4} \cdots R^{\alpha_{2k-1}\alpha_{2k}}{}_{\beta_{2k-1}\beta_{2k}} A^\rho A_\sigma\cr
&&-\ft{k}{2^k} g^{\phantom{XX}}_{\beta_1(\mu}\delta^{\beta_1\beta_2\cdots\beta_{2k}\,\sigma}_{\nu)\alpha_2\cdots
\alpha_{2k}\,\rho}\, R^{\alpha_3\alpha_4}{}_{\beta_3\beta_4}\cdots
R^{\alpha_{2k-1}\alpha_{2k}}{}_{\beta_{2k-1}\beta_{2k}} \nabla^{\alpha_2}\nabla_{\beta_2}
(A^\rho A_\sigma)\,.\label{generalkeom}
\eea
The salient point of the theory (\ref{totalgenlag}) is that in the equations of motion, as in Lovelock gravities, only the Riemann tesnor rather than its derivatives appears.  This implies that the linearized equations involve at most two derivatives.  This is a good indication that the theory can be ghost free. Of course, as in Lovelock gravities, the absence of all ghost modes requires some appropriate choices of coupling constants.  We shall use a concrete example to address this issue in section \ref{cosmo}.

Before ending this section, we would like to comment the connection of the $G^{(k)}$ series with Horndeski gravity, which is given by the series \cite{Horndeski:1974wa}
\be
H^{(k)} = E^{(k)}_{\mu\nu} \nabla^\mu \chi \nabla^\nu \chi\,,
\ee
where $\chi$ is an axion-like scalar.  Horndeski gravity is invariant under the global constant shift of the axion $\chi$.  We may make this symmetry local by introducing a vector with the replacement
\be
\partial_\mu \chi \rightarrow \partial_\mu \chi + A_\mu\,,
\ee
which is invariant under
\be
\chi\rightarrow \chi +\lambda\,,\qquad A\rightarrow A-d\lambda\,.\label{gauge}
\ee
The scalar $\chi$ is then ``eaten'' by $A$ so that the $H^{(k)}$ series becomes our $G^{(k)}$ series.  Thus our $G^{(k)}$ series is the consequence of the gauging of the global shifting symmetry of Horndeski gravity.  The kinetic term for $A$ can be built from the field strength $F=dA$ which is invariant under (\ref{gauge}).  The extra scalar mode in our vector field is a consequence of the violation of the $U(1)$ gauge symmetry.  It implies that we can use the vector as an inflaton to study cosmology, which will be carried out in section \ref{cosmo}.

\section{Gravity with linear curvature coupled to bilinear vector }

In the previous section, we constructed general Einstein-vector gravities involving arbitrary orders of curvature polynomials.  The general theory is complicated and we shall focus on the low-lying examples.  The most general Lagrangian with at most linear curvature terms in general $D$ dimensions is
\be
{\cal L}=\sqrt{-g} \Big( R - 2\Lambda_0 - \ft14 F^2 -\ft12 \mu_0^2 A^2 + \beta R A^2 + \gamma G_{\mu\nu} A^\mu A^\nu\Big)\,.\label{genlagAA}
\ee
Here $(\Lambda_0, \mu_0, \beta, \gamma)$ are constants. The full set of equations of motion are
\bea
&&\nabla_\mu F^{\mu\nu} = \mu_0^2 A^\nu - 2 \beta\,R\, A^\nu - 2\gamma\, A_\mu G^{\mu\nu}\,,\cr
&&G_{\mu\nu}= -\Lambda_0 g_{\mu\nu} + \ft12 (F_{\mu\nu}^2 - \ft14 g_{\mu\nu} F^2) +\ft12\mu_0^2(A_\mu A_\nu - \ft12 g_{\mu\nu}A^2) +\beta\,Y_{\mu\nu} + \gamma\, Z_{\mu\nu}\,,
\eea
where
\bea
Y_{\mu\nu} &=&-R A_\mu A_\nu - G_{\mu\nu} A^2 + (\nabla_\mu\nabla_\nu - g_{\mu\nu}\Box)A^2\,,\cr
Z_{\mu\nu} &=& \ft12 A^2 R_{\mu\nu} + \ft12 R\, A_{\mu} A_\nu -2A^\alpha R_{\alpha (\mu} A_{\nu)} -\ft12 \nabla_\mu\nabla_\nu A^2 + \nabla_\alpha \nabla_{(\mu} \big(A_{\nu)} A^\alpha\big)\cr
&&-\ft12 \Box (A_\mu A_\nu) + \ft12 g_{\mu\nu} \big(G_{\alpha\beta} A^\alpha A^\beta +
\Box A^2 - \nabla_\alpha\nabla_\beta (A^\alpha A^\beta)\big)\,.
\eea
Note that it is necessary to use the identity
\be
[\nabla_{\mu}, \nabla_\nu]A^\rho=R^{\rho}{}_{\sigma\mu\nu} A^\sigma\,,
\ee
to compare the structures in (\ref{generalkeom}) to the above Einstein equations 
of motion.  It is of interest to note that the $\beta$ term resembles the non-minimal coupling in the Brans-Dicke scalar theory \cite{Brans:1961sx}. The cosmological implication of such a Lagrangian were studied in \cite{Jimenez:2009ai,BeltranJimenez:2010uh}.  It was recently demonstrated that
the non-minimal coupling to the Einstein tensor can arise naturally within the context of quadratic curvature terms in Weyl geometry \cite{Jimenez:2014rna} and also in more general geometries introduced in \cite{Jimenez:2015fva}.

The general theory (\ref{genlagAA}) admits solutions with
\be
G_{\mu\nu}=-\Lambda_0 g_{\mu\nu}\,,\qquad A=0\,.
\ee
Thus for a positive, zero or negative bared cosmological constant $\Lambda_0$, the vacuum of the theory is dS, Minkowski or AdS spacetimes respectively. The linear fluctuations of such a vacuum are described by a (massless) graviton and a Proca field with mass
\be
\mu_{\rm eff}^2 = \mu_0^2 -\ft{4D}{D-2} \beta\Lambda_0 + 2 \gamma \Lambda_0\,,\label{mueff}
\ee
Note that even if we start with the vanishing bared mass $\mu_0$, effective Proca mass can be generated in the vacuum, and it is of the order of the cosmological constant. The $U(1)$ gauge symmetry can emerge at the linear order when the parameters are such that $\mu_{\rm eff}=0$.  In this case, the vector field becomes a Maxwell field at the linear order. This statement is true for any spacetime backgrounds described by the Einstein-metrics with a cosmological constant $\Lambda_0$.

We may also consider additional minimally-coupled matter with
\be
{\cal L}_{\rm mat}={\cal L}_{\rm mat} (g_{\mu\nu}, \phi, \nabla_\mu\phi)\,,
\ee
where $\phi$ denotes a generic matter field. In this case, it is advantageous to set $\beta =0$.  The backgrounds with $A=0$ are determined by
\be
G_{\mu\nu}=-\Lambda_0 g_{\mu\nu} + T^{\rm mat}_{\mu\nu}\,.
\ee
In this background, the gauge symmetry of $A$ is broken both by the global effective mass associated with the cosmological constant $\Lambda_0$, but also by the matter energy-momentum tensor $T^{\rm mat}_{\mu\nu}$ which is local and position dependent.

To classify the effects of the cosmological constant and matter energy-momentum tensor to the gauge symmetry, we now consider the following special cases.  The first is
\be
{\cal L}_1=\sqrt{-g} \Big( R - \ft14 F^2 + \gamma G_{\mu\nu} A^\mu A^\nu\Big)\,.\label{lag1}
\ee
In this theory, the maximally-symmetric vacuum is the Minkowski spacetime.  At the linear perturbation level, the gauge symmetry is restored with any Ricci-flat metric, including the Schwarzschild and Kerr black holes.  In a background with distributed matter of energy-momentum tensor $\bar T^{\rm mat}_{\mu\nu}$, the gauge symmetry is broken and the effective Lagrangian for $A$ is
\be
{\cal L}_A=\sqrt{-\bar g} \big(-\ft14 F^2 + \gamma \bar T_{\rm mat}^{\mu\nu} A_\mu A_\nu\big)\,.\label{bilinearA1}
\ee
Here the barred quantities are the background fields. Thus we see an effective Lorentz violation term is introduced to the Maxwell theory and the violation is local and position dependent.  Note that we find some special exact black hole solutions of the theory (\ref{lag1}) in section \ref{bhexact}.

The second case is to introduce a cosmological constant, namely
\be
{\cal L}_2=\sqrt{-g} \Big(R -2\Lambda_0- \ft14 F^2 + \gamma (G_{\mu\nu}+\Lambda_0 g_{\mu\nu}) A^\mu A^\nu\Big)\,.\label{lag2}
\ee
This corresponds to setting $\beta=0$ and $\mu_0^2=-2\gamma\Lambda_0$.
The maximally-symmetric vacuum is the (A)dS spacetime. The gauge symmetry is restored in any Einstein metric of the cosmological constant $\Lambda_0$.  In the background with $A=0$, the linear equation for $A$ is also governed by (\ref{bilinearA1}).  As we shall see later that exact wormholes can be constructed in this theory.

The third special case is
\be
{\cal L}_3=\sqrt{-g} \Big( R -2\Lambda_0- \ft14 F^2 + \gamma\,G_{\mu\nu} A^\mu A^\nu\Big)\,.
\label{lag3}
\ee
The vacuum is also (A)dS. In backgrounds with additional matter but with $A=0$, the linear equation for $A$ is now given by
\be
{\cal L}_A=\sqrt{-\bar g} \big(-\ft14 F^2 -\gamma \Lambda_0 A^2 + \gamma \bar T_{\rm mat}^{\mu\nu} A_\mu A_\nu\big)\,.\label{bilinearA2}
\ee
Thus we see that the gauge symmetry of the Maxwell field is broken by a global Lorentz-preserving mass term induced by the cosmological constant and by Lorentz-violating term associated with local matter energy-momentum tensor.

In the current stage of our Universe, gravity is in general weak and curvature is small, except locally. Furthermore, the energy-momentum tensor of the electric-magnetic field is even weaker and can be analysed at the linear order.  In our theory, the emerging $U(1)$ gauge symmetry may be broken by a global mass, which is of the order of the cosmological constant.  The gauge symmetry can also be broken by the local matter energy-momentum tensor.  In the small region such as the solar system, matter are extremely localized and the space between the Sun and planets can be effectively treated as vacuum.  Of course, if we consider higher order terms such as $G^{(2)}$, local Lorentz-violation can also occur, but its magnitude is of order Riemann tensor squared, and is unlikely to be detectable.  In the large scale such as galaxies and clusters, the energy-density of dark matter are roughly uniformly distributed.  The effect of the local Lorentz-violation term to the light propagation may not be ignored.  This is over and above the effect of curved spacetime on the geodesic motion of light.

In the above discussion, it is crucial that $A$ vanishes as a background field.  This requirement is too strong from the cosmology point of view. As we shall discuss in section \ref{cosmo} that the vector $A$ can also be time dependent and used as an inflaton to construct cosmological solutions.  A non-trivial requirement then arises that $A$ must vanish in the later time of the cosmological evolution so that the $U(1)$ gauge symmetry can emerge at the linear level in later times.  Such time-dependent solutions indeed exist and thus the requirement that $A$ vanishes in a background can be achieved dynamically.

\section{Black hole solutions}

As we mentioned earlier, Einstein metrics are the vacuum solutions of our Einstein-vector theories, it follows that Schwarzschild and Kerr black holes are solutions.  The existence of such solutions implies that Newton's gravity can be recovered in the weak field limit.  On the other hand, the vector $A$ is not a Maxwell field and hence the Reissner-Nordstr\o m (RN) black hole is not a solution.  In this section, we would like to construct solutions carrying the vector hair.  We shall focus our discussion in four dimensions.  The static and spherically-symmetric ansatz is
\be
ds^2 = -h(r) dt^2 + \fft{dr^2}{f(r)} + r^2 d\Omega_{2,\epsilon}^2\,,\qquad A=\phi(r) dt\,,\label{genans}
\ee
where $d\Omega_2^2$ is the metric for the unit round $S^2$, 2-torus or hyperbolic 2-space for $\epsilon=1,0,-1$ respectively.

\subsection{An exact solution}
\label{bhexact}

We do not expect exact solutions for general parameters.  However, for asymptotically-flat spacetime associated with the theory (\ref{lag1}), we find an exact solution when $\gamma=\ft14$, given by
\be
ds^2=-f dt^2 + \fft{dr^2}{f} + r^2 d\Omega^2\,,\qquad A=2\sqrt2\, f\, dt\,,\qquad
f=1 - \sqrt{\fft{r_0}{r}}\,.\label{d4bhsol}
\ee
The solution describes a black hole with the event horizon located at $r=r_0$.  Some low-lying curvature polynomials are given by
\be
R=\fft{3}{4r^2} \sqrt{\fft{r_0}{r}}\,,\qquad
R^{\mu\nu}R_{\mu\nu}=\fft{17 r_0}{32 r^5}\,,\qquad
R^{\mu\nu\rho\sigma}R_{\mu\nu\rho\sigma}=\fft{89r_0}{16r^5}\,.
\ee
Thus we see that the only spacetime singularity $r=0$ is shielded by the event horizon.  The temperature is given by
\be
T=\fft{f'(r_0)}{4\pi} = \fft{1}{8\pi r_0}\,.
\ee
Since the local diffeomorphism-invariant $A_a=E_a^\mu A_\mu$ vanishes on the horizon, where $E_a^\mu$ is the inverse vielbein, we expect that the standard Wald entropy formula holds, giving rise to
\be
S=\pi r_0^2\,.
\ee
This is very different from the back holes in Horndeski gravity \cite{aco} where the Wald entropy formula is not valid owing to the fact that $E_a^\mu \partial_\mu \chi$ does not vanish on the horizon \cite{Feng:2015oea}. The completion of the first law $dM=TdS$ implies that
\be
M=\ft14 r_0\,.
\ee
In other words, the black hole radius is four times of its mass, rather than twice the mass in the Schwarzschild black hole.  Thus the new black hole appears to be more smeared out. In the weak-field limit to recover the Newtonian concept of gravitational force, it is stronger with the ($1/r^{3/2}$)-law than the usual ($1/r^2$)-law associated with the Schwarzschild black hole.  It should be emphasized here that this does not mean that Newton's gravity cannot be recovered.  In fact as mentioned earlier Schwarzschild and Kerr black holes are already solutions of the theory. Rather it predicts existence of additional black hole objects with vector hair that has different powers than the Newton's inverse-squared law.  The existence of such black holes cannot not be ruled out by observational data before we have better understanding how dark matter modifies Newtonian gravity.

\subsection{Discussion on general solutions}

The exact solution (\ref{d4bhsol}) contains only one parameter, rather than two independent parameters, as one expect from having additional vector hair.  The falloff behavior indicate that this is the condensate of the vector modes, rather than the graviton modes, which would yield $1/r$.  The general solution involving two parameters can be established by numerical analysis.  Let us still focus on the theory (\ref{lag1}), the general asymptotic expansions at large $r$ for $(h,f,\phi)$ are
\bea
h &=& 1 + \fft{h_1}{r} + \fft{h_2}{r^2} + \fft{h_3}{r^3} +\cdots \,,\cr
f &=&1 + \fft{f_1}{r} + \fft{f_2}{r^2} + \fft{f_3}{r^3}+ \cdots\,,\cr
\phi &=&\phi_0 + \fft{\phi_1}{r} + \fft{\phi_2}{r^2} + \fft{\phi_3}{r^3} + \cdots\,,\label{asyexpan}
\eea
where all the coefficients can be expressed in terms of three parameters $(\phi_0,\phi_1,f_1)$. For example,
\be
h_1=\fft{2f_1 +\gamma\phi_0(f_1\phi_0 -4\gamma\phi_1)}{2-\gamma \phi_0^2}\,,\quad
f_2=\fft{\phi_1^2}{2(2-\gamma \phi_0^2)}\,,\quad
\phi_2=\fft{\gamma\phi_0\phi_1(\phi_1-f_1\phi_0)}{2(2-\gamma\phi_0^2)}\,,\cdots
\ee
If we assume that a black hole exists with the event horizon located at $r=r_0$, the near-horizon expansions are then
\bea
h &=& \tilde h_1 (r-r_0) + \tilde h_2 (r-r_0)^2 + \tilde h_3 (r-r_0)^3 + \cdots\,,\cr
f &=& \tilde f_1 (r-r_0) + \tilde f_2 (r-r_0)^2 + \tilde f_3 (r-r_0)^3 + \cdots\,,\cr
\phi &=& \tilde \phi_1 (r-r_0) + \tilde \phi_2 (r-r_0)^2 + \tilde \phi_3 (r-r_0)^3 + \cdots\,,
\eea
with
\bea
&&\tilde f_1=\fft{4\tilde h_1}{r_0(4\tilde h_1 + \tilde \phi_1^2 r_0)}\,,\quad
\tilde \phi_2=-\fft{\tilde \phi_1}{r_0} + \fft{\gamma r_0\tilde \phi_1^5}{16\tilde h_1^2}\,,\quad
\tilde h_2 = (\ft14-\gamma)\tilde \phi_1^2 + \fft{\tilde h_1}{r_0} + \fft{\gamma r_0 \tilde\phi_1^4}{16\tilde h_1}\,,\cr
&&\tilde f_2 = -\fft{16\tilde h_1^2 - 4 r_0\tilde h_1 \tilde \phi_1^2 + 3\gamma
r_0^2\tilde\phi_1^4}{4r_0^2\tilde h_1 (4\tilde h_1 + r_0\tilde \phi_1^2)}\,,\qquad
\hbox{etc.}
\eea
Thus the near-horizon expansion is specified by parameters, $(r_0,\tilde h_1,\tilde \phi_1)$.  One of the parameter however is ``trivial'' in that it is associated with the time scaling and it should be fixed such that when integrated out to asymptotic infinity, we have $g_{tt}=-1$.  Thus analogous to the RN black hole, the general black hole solution with vector hair contains two parameters.  The asymptotic three parameters $(\phi_0,\phi_1,h_1)$ are related by one constraint fixed by the horizon condition. In particular, when $\gamma=0$, the solution reduces to the RN black holes.  We have performed the numerical analysis integrating out from the horizon to asymptotic infinity and indeed black holes exist for some appropriate non-vanishing $\gamma$.

The asymptotic expansion (\ref{asyexpan}) is no longer valid when
\be
\gamma \phi_0^2 =2\,.
\ee
In this case, black holes may still exit when $\gamma=1/4$, given by (\ref{d4bhsol}).  This is however not the most general solution. We find that the general asymptotic expansion now becomes
\bea
h &=& 1 - \fft{\phi_1}{\sqrt r} - \fft{m}{r} - \fft{5m^2}{2\phi_1 r^{\fft32}}+\cdots\,,\cr
f &=& 1 -\fft{\phi_1}{\sqrt r} + \fft{m}{r} + \fft{3m^2}{2\phi_1 r^{\fft32}}+\cdots\,,\cr
\phi&=& 2\sqrt2 - \fft{2\sqrt2 \phi_1}{\sqrt r} + \fft{\sqrt2\,m}{r}  + \fft{\sqrt2\phi_1 m}{r^{\fft32}} +
\cdots
\eea
Thus we see that the general solution contain an additional parameter $m$, which corresponds to the condensate of the graviton. However, we can set $m=0$ and obtain the exact solution (\ref{d4bhsol}), but there is no smooth $\phi_1=0$ limit when $m$ is non-zero.

So far we have considered black holes of the theory (\ref{lag1}) and all the solutions are asymptotic to the Minkowski spacetime.  For the more general theory with a cosmological constant, the asymptotic behavior of the vector $A$ can be more complicated.  After failing to find some exact (A)dS black holes with the vector hair, we shall not discuss this further.

\section{Wormhole solutions}
\label{whsol}

\subsection{A superpotential method}

The general static solutions of the ansatz (\ref{genans}) is unlikely analytical.  In this section, we adopt a superpotential method to obtain some special exact solutions.  Such a method is typically used for constructing solutions admitting Killing spinors \cite{Cvetic:2000db}.  Non-supersymmetric solutions such as non-extremal RN-AdS black holes can also be constructed using this method \cite{Lu:2003iv}. Many exact cosmological and domain-walls solutions were also constructed in $f(R)$ gravity using this method \cite{Liu:2011am}.  It turns out that this method works for the Lagrangian (\ref{lag2}), i.e. $\beta=0$ and $\mu_0^2 = -2\gamma \Lambda_0$. New solutions turn out to describe wormholes that connect two symmetric (A)dS or Minkowski spacetimes.

Again we shall first focus on the construction in four dimensions, and present the solutions in general dimensions in the appendix.  We start by following Refs.~\cite{Cvetic:2000db,Lu:2003iv}
and rewrite the general static ansatz (\ref{genans}) as
\be
ds^2=a^2 b^4\,d\rho^2 -a^2 dt^2 + b^2 d\Omega_2^2\,,\qquad A=\phi\, dt\,,
\ee
where $(a,b,\phi)$ are functions of the radial coordinate $\rho$.  We find that the effective one-dimensional Lagrangian reduced from ${\cal L}_2$ in (\ref{lag2}) is $L=T - V$ where the kinetic $T$ and potential $V$ energies are
\bea
T &=& \Big(2 - \fft{\gamma\phi^2}{a^2}\Big) \fft{a'b'}{a b} +
\Big(1 + \fft{\gamma \phi^2}{a^2}\Big) \fft{b'^2}{b^2} +
\fft{2\gamma \phi \phi' b'}{a^2 b} + \fft{\phi'^2}{4a^2}\,,\cr
V &=& \fft14 b^2 (-4 a^2 + 4\Lambda_0 a^2 b^2 - 2 \gamma \phi^2 +2 \gamma \Lambda_0 b^2 \phi^2)\,.
\eea
In this paper, a prime on a function denotes a derivative with respect to the variable of the function, which is $\rho$ in this case. (It is perhaps a misnomer to call $T$ the kinetic energy since the derivative is with respect $\rho$, rather than the time $t$.) The superpotential method is first to treat the kinetic term $T$ as some one-dimensional $\sigma$-model
\be T=\ft12 g_{ij} (X^i)' (X^j)'\,.
\ee
If the potential $V$ can be expressed in terms of a superpotential $W$ as
\be
V=-\ft12 g^{ij} \fft{dW}{dX^i}\fft{dW}{dX^j}\,,
\ee
the Lagrangian then admits special solutions that satisfy the first-order equations
\be
(X^i)'=g^{ij} \fft{\partial W}{\partial X^j}\,.
\ee
For the theory (\ref{lag2}), we find that a superpotential indeed exists, given  by
\be
W=\fft1{a}(2 a^2 + \gamma \phi^2)\sqrt{b(b \,\epsilon-\ft13\Lambda_0b^3 - m)}\,,
\ee
where $m$ is an arbitrary constant, which turns out to be the mass of the solution.
The resulting first-order equations are
\bea
a'=\fft{ab (3m-\Lambda_0 b^3)(2a^2 + \gamma \phi^2)^2}{6W (2a^2 - \gamma\phi^2)}\,,\quad
b' = -\fft{b^2 (3m-\Lambda_0 b^3)(2a^2 + \gamma \phi^2)^2}{3W}\,,\quad \phi'=0\,.
\label{lag2fo}
\eea

\subsection{Asymptotically-flat or (A)dS wormholes}

The set of first order equations (\ref{lag2fo}) can be solved straightforwardly.  In terms of the original ansatz (\ref{genans}), we find
\be
f=-\ft13 \Lambda_0 r^2 +\epsilon - \fft{m}{r}\,,\qquad
h=h_0 + \ft12 \Big(1 + \sqrt{1 + \fft{4h_0}{f}}\Big)f\,,\qquad
\phi=\phi_0\equiv \sqrt{-\fft{2h_0}{\gamma}}\,.
\ee
The solution contains two integration constants, the mass parameter $m$ and the vector constant hair $\phi_0$. When $h_0=0$, the solution reduces to the Schwarzschild-(A)dS black hole.  For $h_0>0$, the solution is a wormhole connecting two (A)dS spacetimes, with the wormhole throat located at $f(r_0)=0$.  The reality condition requires that $\gamma<0$.  Note that for non-vanishing $\Lambda_0$, the value of $\epsilon$ can take values $(1,0,-1)$.

For vanishing $\Lambda_0$, the wormhole is asymptotic to Minkowski spacetime and hence we must set $\epsilon=1$.  Setting the convention that $h=1$ as $r\rightarrow \infty$, the solution has three branches depending on the value of $h_0$
\be
h=\left\{
    \begin{array}{ll}
      h_0 +\ft12 (h_0-1)^2 \Big(1 + \sqrt{1 + \fft{4h_0}{(h_0-1)^2f}}\Big)f, & \qquad\qquad 0<h_0<1;\\
1, &\qquad\qquad h_0=1;\\
      h_0 +\ft12 (h_0-1)^2 \Big(1 - \sqrt{1 + \fft{4h_0}{(h_0-1)^2f}}\Big)f, & \qquad\qquad h_0>1\,.
    \end{array}
  \right.\label{hworm}
\ee
The function $f$ and $\phi$ are the same for all branches, given by
\be
f=1 - \fft{m}{r}\,,\qquad \phi=\phi_0\equiv \sqrt{-\fft{2h_0}{\gamma}}\,,\qquad \gamma<0\,.
\ee
Although we set out to construct special solutions using the superpotential method, the superpotential $W$ we obtained contains already an arbitrary constant $m$ and hence the final solutions all contain two independent parameters.  We expect these are the most general wormhole solutions of the theory.  This is analogous to the situation in \cite{Lu:2003iv} where the most general RN-AdS black holes were constructed using the superpotential method.  The fact that the wormhole solutions can arise from the first-order equations via the superpotential is suggestive that these solutions may be stable.

\section{Lifshitz spacetimes and black holes}

The Lifshitz spacetime was introduced in \cite{Kachru:2008yh}. The simplest theory that admits such a geometry is perhaps the Einstein-Proca theory together with a negative cosmological constant \cite{Taylor:2008tg}.  Our vector field $A$ resembles the Proca field in that besides its kinetic term, it is bare and quadratic in the action with no $U(1)$ gauge symmetry.  We find that indeed our Einstein-vector theory admits Lifshitz vacua, and furthermore exact Lifshitz black holes can be constructed.

\subsection{Lifshitz black holes}

\subsubsection{Lifshitz spacetimes}

The Lifshitz solutions take the form
\be
ds^2 = \ell^2 \big(-r^{2z} dt^2 + \fft{dr^2}{r^2} + r^2 (dx_1^2+ dx_2^2)\big)\,,\qquad
A=q r^z\,\ell dt\,.
\ee
The metric is homogeneous and invariant under the Lifshitz scaling
\be
r\rightarrow \lambda^{-1} r\,,\qquad t\rightarrow \lambda^z\,t\,,\qquad x_i\rightarrow \lambda \,x_i\,.
\ee
The general solution of the theory (\ref{genlagAA}) in four dimensions turns out to be
\be
\ell^2=\fft{2}{\mu_0^2}(3\gamma -2\beta (3+2z + z^2) + z)\,,\qquad
q^2=\fft{2(z-1)}{\gamma (1-z) + 2\beta (z-1) + z}\,,
\ee
where the Lifshitz exponent $z$ is determined by a cubic polynomial equation of $z$:
\be
\Lambda_0=-\fft{q^2}{4\ell^2(z-1)}\big(6(1-z)\gamma + 4\beta(z-1)(3+2z+z^2) + z(4 + z+z^2)\big)\,.
\ee

\subsubsection{Black holes}
\label{lbh}

To construct a black hole, we add a non-extremal factor $f(r)$ to the Lifshitz spacetimes obtained in the previous subsection and consider
\be
ds^2 = \ell^2 \big(-r^{2z} f dt^2 + \fft{dr^2}{r^2} + r^2 (dx_1^2+ dx_2^2)\big)\,,\qquad
A=q r^z\,\ell f dt\,.
\ee
We find that $f$ is given by
\be
f=1 -\big(\fft{r_0}{r}\big)^{1+\fft{1}{2}z}\,,
\ee
with
\bea
q^2 &=& \fft{4(z-1)(4-z)(3z+2)}{z(z+2)(z+10)}\,,\qquad
\ell^2=\fft{z(z+2)(2+5z+4z^2-2z^3)}{\mu_0^2(z-1)(4-z)(3z+2)}\,,\cr
\gamma &=& -\fft{(z+2)(8-8z+5z^2)}{2(z-1)(z-4)(3z+2)}\,,\qquad
\beta = -\fft{z^2-z+2}{2(z-1)(3z+2)}\,,\cr
\Lambda_0 &=& -\fft{z(z+2)(3+5z+z^2)q^2}{4(z-1)(4-z)(3z+2)\ell^2}\,.
\eea
We are particularly interested in solutions with $z>1$.  For $\mu_0^2>0$, the reality condition requires that $1<z<2.96$, whilst for $\mu_0^2<0$, it requires that $2.96<z<4$.  Thus the solutions contain two integers for $z$, namely $z=2$ and 3.  The thermodynamical quantities are
\be
M=\fft{r_0^{z+2}\ell^2}{16\pi}\,,\qquad T=\fft{(z+2) r_0^z}{8\pi}\,,\qquad
S=\ft14 r_0^2 \ell^2\,.
\ee
The first law $dM=TdS$ can be easily established to be satisfied.

In the AdS/CMT correspondence, the $z=2$ Lifshitz black holes are of particular interest and we present the result explicitly
\bea
ds^2 &=& \fft{6}{\mu_0^2}\Big(-r^4 f dt^2 + \fft{dr^2}{r^2 f} + r^2 (dx_1^2 + dx_2^2)\Big)\,,\qquad A=\fft{2r^2}{\mu_0} f\, dt\,,\cr
f &=& 1-\fft{r_0^2}{r^2}\,,\qquad \gamma=-\ft32\,,\qquad \beta=-\ft14\,,\qquad
\Lambda_0=-\ft12\mu_0^2\,.\label{lbhsol}
\eea

\subsubsection{Charged black holes}
\label{clbh}

In the application of the AdS/CMT correspondence, the concept of ``realistic'' field contents in a theory is much vague.  We may couple an additional Maxwell field ${\cal A}$ to the Lagrangian (\ref{genlagAA}), namely
\be
\sqrt{-g} \Big(-\ft14 {\cal F}^2\Big)\,,\label{maxwell}
\ee
where ${\cal F}=d{\cal A}$.  For appropriate parameters, we find a charged $z=6$ Lifshitz black hole. We present the solution in four dimensions
\bea
ds^2 &=& \ell^2 \Big(-r^{12}f dt^2 + \fft{dr^2}{r^2 f} + r^2 (dx_1^2 + dx_2^2)\Big)\,,\qquad
A=\sqrt{\ft{5}{8(1+10\beta)}}\, f\, r^6\,\ell dt\,,\cr
{\cal A} &=& (\psi_0 + Q r^4)\,\ell dt\,,\qquad f=1 - \fft{4(1 + 10\beta)Q^2}{(4 + 25\beta) r^4}\,,\cr
\Lambda_0 &=& \fft{(7 + 40\beta)\mu_0^2}{256\beta (1 + 10\beta)}\,,\qquad
\gamma=-2-30\beta\,,\qquad \ell^2 = -\fft{384\beta}{\mu_0^2}\,.\label{chargedbh}
\eea
where the parameter $-\fft{1}{10}<\beta\le \infty$, with the following constraint
\be
\left\{
  \begin{array}{ll}
    -\fft{1}{10}<\beta<0\,, & \qquad\qquad \mu_0^2>0\,; \\
    \beta=0\,, & \qquad\qquad \mu_0^2=0\,; \\
    \beta>0\,, & \qquad\qquad \mu_0^2<0\,.
  \end{array}
\right.
\ee
Note that we can a smooth limit $(\mu,\beta)\rightarrow 0$, with $-384\beta/\mu^2\equiv \ell^2$ kept fixed, the solution is then to the equations from Lagrangian (\ref{lag2}) augmented with (\ref{maxwell}).  The thermodynamical quantities of the black holes are
\be
M=\fft{15\beta r_0^8}{64\pi(1+10\beta)}\,,\qquad Q_e=\fft{Q}{4\pi}\,,\qquad \Phi_e=-Q r_0^4\,,\qquad
S=\ft14 r_0^2\,,\qquad T=\fft{r_0^6}{\pi}\,.
\ee
Note that $M$ vanishes when $\beta=0$. The first law of thermodynamics $dM=TdS + \Phi_e dQ_e$ is then straightforwardly satisfied.  The generalized Smarr relation associated with the scaling symmetry of the solution \cite{Liu:2015tqa}, namely
\be
M=\ft14 (TS + \Phi_e Q_e)\,,
\ee
is satisfied. Note that the electric potential of ${\cal A}$ is divergent and hence $\Phi_e$ is not the difference between its values at infinity and on the horizon.  This quantity was derived adopting the Wald formalism for Lifshitz black holes in \cite{Liu:2014dva}. Note the generalizations of these black hole solutions to arbitrary dimensions can be found in the appendix.

It is worth commenting again that in the charged black hole solution, we can set the parameter $\beta=0$, whilst in the neutral black hole in the previous subsection, the parameter $\beta$ is non-vanishing.

\subsection{Viscosity/entropy ratio}

Lifshitz black holes can be viewed as generalizations to AdS planar black holes. There is an $SO(2)$-rotational symmetry between the coordinates $x_1$ and $x_2$.  The geometry can be viewed as the gravitational dual to some ideal fluid, and the transverse and traceless mode can be used to calculate the shear viscosity in the boundary field theory \cite{KSS0}.  We consider such a perturbation by making the replacement of the planar section of the black holes
\be
dx^i dx^i \longrightarrow dx^i dx^i + 2\Psi(r,t)\, dx^1 dx^2\,,\label{pert}
\ee
For the Lifshitz black hole (\ref{lbhsol}), we find that the
mode $\Psi(r,t)$ satisfies the linearised equation
\bea
0&=&2rzf^2 \big((7z^2 - 8z +4) f^2 - (z+10)(z+2)\big) \ddot\Psi\cr
 &&+ zf\big(2(7z^2-8z+4) f^2 + (z+2)(13z^2-30z-32) f - (z+10)(z+2)^2\big)\dot \Psi\cr
&&\ft2{r^{2z+1}}\big((3z^3+12z^2-20z+32)f + z(z+2)(z+10)\big) \Psi''\,.
\eea
For an infalling wave which is purely ingoing on the horizon, the solution
for a low-frequency wave is given by
\be
\Psi=e^{-{\rm i} \omega t} e^{-\fft{{\rm i}\omega}{4\pi T} \log f}\big(1-{\rm i}\omega U(r)\big) +
{\cal O}(\omega^2)\,,
\ee
where $r_0$ is the horizon radius, $T$ is the temperature and
\be
U' =- \fft{2(z-4)(3z+2) r_0^{\fft12(2-z)}}{r\big((z-4)(3z+2) r^{\fft12(z+2)} - (7z^2-8z+4)
r_0^{\fft12(z+2)}\big)}\,.
\ee
The viscosity can be determined, using the procedure in \cite{KSS0} and we find
\be
\eta = \fft{(z+2) r_0^{z+2}\ell^2}{128\pi^2} = \fft{r_0^2\ell^2}{16\pi}=\fft{S}{4\pi}\,.
\ee
Thus
\be
\fft{\eta}{S}=\fft{1}{4\pi}\,.
\ee
The same conclusion can be drawn for the charged Lifshitz black holes (\ref{chargedbh}).  Thus there is no surprise in this ratio, unlike the situation in Horndeski gravity where the ratio can be smaller than $1/(4\pi)$ \cite{Feng:2015oea}.

\section{Cosmological solutions}
\label{cosmo}

In this section, we consider cosmological solutions of the general theory (\ref{genlagAA}) in four dimensions.  Typically a vector such as Maxwell field has a tendency of breaking the homogeneous and isotropic cosmological principle. However, in our theory, the field $A$ contains also a longitudinal scalar mode in additional to the transverse modes. We can thus use the scalar mode within $A$ as an inflaton to construct cosmological models. We consider FLRW-type homogeneous and isotropic ansatz
\be
ds^2 = -dt^2 + a(t)^2 (dx_1^2 + dx_2^2 + dx_3^2)\,,\qquad A=\phi(t)\, dt\,.\label{cosmoans}
\ee
If we would like also to use the theory to describe a universe with emergent $U(1)$ gauge symmetry, we must then also require that $\phi\rightarrow 0$ as $t\rightarrow \infty$. With this condition satisfied, the background with $A=0$ becomes a consequence of dynamical evolution.  Of course, one may also use the theory to study the inflation only, without being concerned with gauge symmetry, the restriction of $A$ can be relaxed.

\subsection{de Sitter universe}

There are two types of de Sitter spacetimes. One is the maximally-symmetric de Sitter vacuum, with $A=0$.  The scaling factor is
\be
a=e^{\sqrt{\fft{\Lambda_0}{3}}\, t}\,.
\ee
The other type has non-vanishing $A$. The scaling factor $a$ and the effective cosmological constant are
\be
a=e^{\lambda t}\,,\qquad \Lambda_{\rm eff}=3\lambda^2\,,\qquad \lambda^2=\fft{\mu_0^2}{6(4\beta-\gamma)}\,.
\ee
The solution for $\phi$ depends on the values of the parameters. For $\beta<\gamma$, it is given by
\be
\phi=\fft{\sqrt{
\Lambda_{\rm eff}-\Lambda_0+e^{-\fft{\gamma-\beta}{\beta} \lambda t}}}{\sqrt{(\gamma-\beta)\Lambda_{\rm eff}}}\,.
\ee
The maximum symmetry of the de Sitter metric is broken by the vector field, although the homogeneity and isotropy of the spatial section are preserved.
The reality condition throughout the time $t\in (-\infty,\infty)$ requires that the effective cosmological constant is no less than the bared cosmological constant, i.e.~$\Lambda_{\rm eff}\ge \Lambda_0$.  Note that we can set $\Lambda_0=0$, which implies that the universe can expand with an acceleration without a cosmological constant.  Thus our theory provides a new candidate for dark energy.  Interestingly, the Hubble constant $\lambda$ can be made arbitrary small with the parameter $\mu_0$. However, this model does not resolve the cosmological constant problem; rather it turns a small cosmological constant to the small $\mu_0$.
Note as $t$ runs to infinity, for $\beta>0$, $\phi$ approaches some finite constant, that vanishes as $\Lambda_{\rm eff}=\Lambda_0$.

For $\beta>\gamma$, it is more natural to take $\phi$ to be
\be
\phi=\fft{\sqrt{e^{\fft{\beta-\gamma}{\beta} \lambda t}
-\Lambda_0+\Lambda_{\rm eff}}}{\sqrt{(\beta-\gamma)\Lambda_{\rm eff}}}\,.
\ee
The reality condition throughout the time $t\in (-\infty,\infty)$ requires that $\Lambda_{\rm eff} \le \Lambda_0$. As $t$ runs to infinity, $\phi$ also diverges if $\beta>0$ and it converges if $\beta<0$.  There are two more special cases:
\bea
\beta=0:\qquad\phi=\sqrt{\fft{\Lambda_{\rm eff}-\Lambda_0}{\gamma\Lambda_{\rm eff}}}\,;\qquad
\beta=\gamma:\qquad
\phi=\fft{\sqrt{(\Lambda_{\rm eff}-\Lambda_0)\,t}}{\sqrt{\gamma\,\Lambda_{\rm eff}}}\,.
\eea
In all these inflating solutions, with or without a cosmological constant, the initial cosmic singularity problem persists, by the same argument presented in \cite{Borde:2001nh}.  However, our theory also admits bounce solutions that may resolve the issue, which we shall discuss next.

\subsection{de Sitter bounce}

The equations associated with the cosmological ansatz (\ref{cosmoans}) turn out to be solvable completely. For simplifying the presentation, we rewrite the parameters in terms of $(\mu,\nu)$, defined by
\be
\mu_0^2=12\beta\, \mu^2\nu\,,\qquad \gamma=\fft{2\beta(2\nu-1)}{\nu}\,.
\ee
The general solution is given by
\be
a=[\cosh(\mu t)]^\nu\,,\qquad \phi^2=\sinh(\mu t) [\cosh(\mu t)]^{1-3\nu}\, \psi\,,
\ee
where
\be
\dot \psi = \fft{\mu\nu}{\beta} [\cosh(\mu t)]^{3\nu-2} -
\fft{\Lambda_0 [\cosh(\mu t)]^{3\nu}}{3\beta\mu\nu\, [\sinh(\mu t)]^2}\,.
\ee
This can be solved in terms of hypergeometric functions, given by
\bea
\psi &=&\psi_0+ \fft{\nu}{\beta}\,{}_2F_1[\ft12, -\ft32(\nu-1); \ft32; -\sinh^2(\mu t)] \sinh(\mu t)\cr
&& + \fft{\Lambda_0}{3\beta\mu^2\nu\, \sinh \mu t}\, {}_2F_1[-\ft12, -\ft12(3\nu-1); \ft12; -\sinh^2(\mu t)]\,,
\eea
where $\psi_0$ is an integration constant. The metric function $a$ clearly describes a bounce at $t=0$. Since inflation itself does not resolve the initial cosmic singularity \cite{Borde:2001nh}, a bounce solution is of particular interest.  The reality condition for $\phi$ for all the comoving time range $(-\infty,\infty)$ gives some restrictions on $\Lambda_0$ and $\psi_0$. We shall focus on the case with $\beta\nu>0$.  First it is necessary that we must have $\Lambda_0\ge 0$. For $\Lambda_0=0$, the integration constant $\psi_0$ must vanish, whilst for $\Lambda_0>0$, $|\psi_0|$ must be less than some critical values depending on $\Lambda_0$.  With the reality condition satisfied, the cosmological solutions describe a class of smooth de Sitter bounce universes with the effective cosmological constant
\be
\Lambda_{\rm eff}=3\mu^2\nu^2\,,
\ee
as $t\rightarrow \pm \infty$.  For $\nu\le \fft23$, the function $\phi^2$ is non-negative and diverges as $t\rightarrow \pm \infty$, provided that $\Lambda_0\ge 0$.

For $\nu>\fft23$, in the asymptotic $t\rightarrow \pm \infty$ region, the function $\phi$ approaches a constant, given by
\be
\phi^2 \rightarrow \fft{\nu}{(3\nu-2)\beta} \Big(1 - \fft{\Lambda_0}{\Lambda_{\rm eff}}\Big)\,.
\ee
Thus the full reality condition requires that $0\le \Lambda_0\le\Lambda_{\rm eff}$.  In one limit, $\Lambda_0=0$, we have
\be
\phi^2 = \fft{\nu}{\beta} [\cosh(\mu t)]^{1-3\nu} \sinh^2(\mu t)
\,{}_2F_1[\ft12, -\ft32(\nu-1); \ft32; -\sinh^2(\mu t)]\,.
\ee
This de Sitter bounce is generated by ``dark energy'' without bared cosmological constant. Although the hypergeometric function is already rather straightforward, we present the simplest $\nu=1$ solution in which the hypergeometric function becomes identity.  The $\nu=1$ bouncing universe is
\be
ds^2= -dt^2 + \cosh^2(\mu t) (dx_1^2 + dx_2^2 + dx_3^2)\,,\qquad
A=\fft{\tanh \mu t}{\sqrt{\beta}}\, dt\,.\label{bounce1}
\ee
In the other limit with $\Lambda_0=\Lambda_{\rm eff}$, we have
\be
\phi^2 = \fft{\nu}{\beta} [\cosh(\mu t)]^{1-3\nu}\,{}_2F_1[\ft12, -\ft32(\nu-1); \ft12; -\sinh^2(\mu t)]\,.
\ee
The $\nu=1$ solution is given by
\be
ds^2= -dt^2 + \cosh^2(\mu t) (dx_1^2 + dx_2^2 + dx_3^2)\,,\qquad
A=\fft{1}{\sqrt{\beta}\,\cosh \mu t}\, dt\,.\label{bounce2}
\ee
The $\Lambda_{\rm eff}=\Lambda_0$ solutions are particularly interesting, since as $t\rightarrow \pm \infty$, we have $\phi\rightarrow 0$.  Furthermore, it turns out in this case, the effective mass of the ``photon'' field, defined in (\ref{mueff}), vanishes precisely. Thus the solutions describe the bounce between two de Sitter vacua with $A=0$, whose linear spectrum contains precisely one graviton and one photon.

\subsection{The issue of ghosts}

The existence of the cosmological bouncing is rather surprising, and one is entitled to suspect whether the solutions involve ghost excitations.
As we have established in section 2 that our Einstein-vector theory is absent from ghosts associated with the higher derivative terms in the linearized limit.  However, ghosts may still arise if the theory contains a wrong sign for the kinetic terms of the two derivative theory.  Such a possibility exists even in Lovelock gravity. Owing to the term $G_{\mu\nu} A^{\mu} A^\nu$, the issue is not easy to establish.  It is clear from the construction of the theory that the transverse modes of the vector are not ghost like. We then consider a traceless and transverse tensorial perturbation, namely
\be
dx_1^2 + dx_2^2 + dx_3^2 \rightarrow dx_1^2 + dx_2^2 + dx_3^2 + 2 \psi(t) dx_1 dx_2\,.
\ee
This mode does not couple with other linear excitations. The Lagrangian for $\psi$ takes form
\be
L_\psi = \ft12 a^3 K_1 (\dot \psi + K_2 \psi)^2 + K_3 \psi^2\,,
\ee
where $K_1,K_2$ and $K_3$ are functions of $(a,\phi)$ and their time derivatives.  The crucial quantity for determining whether the excitation is ghost or not is $K_1$, given by
\be
K_1 = 1 +\fft{\beta (1-3\nu)}{\nu}\phi^2\,.\label{K1}
\ee
Note that if we set $\beta=0=\gamma$, for which $K_1=1$, the theory corresponds to the Einstein gravity for which there is no ghost excitation.  Thus for the parameter rang $0<\nu\le \ft13$, $K_1$ is non-negative and hence graviton mode $\psi$ is not a ghost.  Even for $\nu>\ft13$, one can adjust the parameters such that $\psi$'s kinetic energy is positive in all $t>t_0$ for some $t_0>0$. For example, when $\Lambda_{\rm eff}=\Lambda_0$, $\phi^2\rightarrow 0$ as $t\rightarrow \infty$.  This implies that $K_1$ must become positive as $t$ increases, since $K_1=1$ for $t\rightarrow \infty$.  It should be remarked that in cosmological evolution, one is concerned with the late time instability.  It is perfectly reasonable to have ghost excitation during the bounce as long as the ghost disappears in late time after the bounce.  This exactly happens with the $\psi$ mode for $\nu>\ft13$, for which the kinetic energy for $\psi$ becomes positive as $t\rightarrow \pm \infty$.

The situation is subtler for the scalar modes, since the scalar components of the metric and vector couple.  The analysis however can be performed at the full nonlinear level. The reduced one-dimensional Lagrangian is given by
\be
L=-\ft12g_{ij} \dot X^i \dot X^j + \ft12\mu^2 a^3 \phi^2\,,
\ee
where $X^i=(a,\phi)$ and
\bea
g_{ij} &=&
\left(
  \begin{array}{cc}
    K_1 & \tilde K_2 \\
    \tilde K_2 & 0 \\
  \end{array}
\right)\,,\qquad \tilde K_2=-12\beta a^2\phi\,,
\eea
where $K_1$ is given by (\ref{K1}).  At the first sight, the off-diagonal term associated with $\tilde K_2$ seems to suggest that there is an inevitable ghost since the two eigenvalues of $g_{ij}$ must have opposite signs.  However, this not a problem since one is associated with gravity and another is associated with the matter.  For example the same situation arises in Einstein gravity with a minimally-coupled scalar, i.e ${\sqrt{-g}} (R - \ft12 (\partial\varphi)^2)$.  The theory has no ghost but kinetic terms of $a$ and $\varphi$ have the opposite sign.  The sign of the $K_1$ factor however matters, as we have already analysed for the tensor perturbation earlier.

\section{Domain Walls and Randall-Sundrum II}
\label{sec:dw}

\subsection{Exact solutions}

In this section, we consider domain-wall solutions and obtain metrics that realise the Randall-Sundrum II (RSII) scenario \cite{Randall:1999vf}.  For this purpose, we consider the general Lagrangian (\ref{genlagAA}) in five dimensions.  The ansatz is
\be
ds^2=dr^2 + a(r)^2 (-dt^2 + dx_1^2 + dx_2^2 + dx_3^2)\,,\qquad A=\phi(r)\, dr\,.\label{dwans}
\ee
We rewrite the parameters in terms of $(\mu, \nu)$ given by
\be
\mu_0^2=16\beta\, \mu^2\nu\,,\qquad \gamma=\fft{2\beta(5\nu+2)}{3\nu}\,.
\ee
The domain-wall solutions are given by
\bea
a &=& \fft{1}{\cosh^\nu(\mu r)}\,,\qquad \phi^2=\fft{3\nu}{2\beta} [\cosh(\mu r)]^{1 + 4\nu}\psi\,,\cr
\psi &=& \psi_0 \sinh(\mu r) +\sinh^2(\mu r)\, {}_2F_1[\ft12, \ft32+2\nu; \ft32; -\sinh^2(\mu r)]\cr
 &&+ \fft{\Lambda_0}{\Lambda_{\rm eff}}\, {}_2F_1[-\ft12,\ft12+2\nu;\ft12;-\sinh^2(\mu r)]\,,
\eea
where $\Lambda_{\rm eff}=-6\mu^2\nu^2$, and $\psi_0$ is an integration constant.  Keep in mind that the parameters of the solution should be that $\phi^2$ is non-negative for all $r\in (-\infty,\infty)$.

The metric function of the domain-wall is characterized by the sign of the parameter $\nu$. For positive $\nu$, the metric behaves like the RSII scenario \cite{Randall:1999vf}.  When $\nu$ is negative, the metric is a wormhole (or more precisely brane) connecting two Minkowski boundaries of AdS spacetimes, with effective cosmological constant $\Lambda_{\rm eff}$.

\subsection{Trapping of gravity?}

In the previous subsection, we obtain non-trivial domain-wall solutions of the RSII scenario. It is of interest to study whether the domain walls can trap gravity or not.  For simplicity, we consider the solutions with $\Lambda_0=0$.  We follow the general procedure outlined in \cite{Randall:1999vf} and begin by considering the traceless and transverse tensorial perturbation in the brane $dx^\mu dx^\nu\eta_{\mu\nu}$. The linearized equation of this tensor mode has the same form as that of a scalar $\Psi(r,x^\mu)$ satisfying
\be
\Big(\Box_5 + P_1 \fft{d}{dr} + P_2 \fft{d^2}{dr^2}\Big)\Psi=0\,,\label{rspert}
\ee
where $\Box_5$ is the Laplacian of the $D=5$ metric and
\bea
P_1 &=&-\fft{(2 \beta + \gamma) \mu_0^2 a^2 \phi^2 +
 4 \big(6 (2 \beta + \gamma) + (12 \beta^2 - 20 \beta \gamma + 3 \gamma^2) \phi^2\big) a'^2}{8\beta (2 + (2\beta-\gamma)\phi^2) (a^2)'}\,,\cr
P_2 &=& \fft{2\gamma\phi^2}{2 + (2\beta-\gamma)\phi^2}\,.
\eea
Analogous extra $P_1$ and $P_2$ terms in (\ref{rspert}) were also seen in $f(R)$ gravity \cite{Zhong:2010ae}, and some exact domain wall examples can be found in \cite{Liu:2011am}. In order to study the behavior of the $\Psi$, it is advantageous to write the domain wall in terms of the conformal frame, namely
\be
ds^2 = \fft{1}{b^2} (dz^2 + dx_1^2 + dx_2^2 + dx_3^2 -dt^2)\,,\qquad
A=\tilde\phi\, dz\,,
\ee
This implies that
\be
a=\fft{1}{b}\,,\qquad dr=\fft{dz}{b}\,,\qquad \phi=b\,\tilde\phi\,.
\ee
We now perform the Fourier expansion in the momentum space and write
\be
\Psi=\psi(z)\chi(z) e^{{\rm i} p\cdot x}\,,
\ee
where $\chi$ satisfies
\be
\fft{\chi'}{\chi}= -\fft{3\nu(\nu+1)b'^2 + 2\beta (4\nu+1) b^2 \tilde\phi^2(\mu^2\nu^2 +
(\nu+1)b'^2)}{\nu\big(3\nu + \beta (4\nu+1)b^2\tilde \phi^2\big)(b^2)'}\,.
\ee
The function $\psi(z)$ then satisfies a Schr\"odinger equation
\be
(-\fft{d^2}{dz^2} + V)\psi(z) = m^2 \psi(z)\,,\qquad m^2\equiv p_\mu p^\mu\,,\label{schr}
\ee
with the potential $V$ may also depend on $m^2$ as well.  We are interested in the potential for the effective massless gravity in the brane, corresponding to $m^2=0$; it is given by
\bea
V &=& \fft{1}{2[\nu\big(3\nu + 2\beta (4\nu+1)b^2\tilde \phi^2\big)(b^2)']^2}\Big\{
-4 \mu^4 \nu^4 \beta( 4\nu+1)^2 b^4  \tilde\phi^4\cr
&&+2 \mu^2 \nu^2 \big(3 \nu +
   2 \beta (4 \nu+1) b^2 \tilde \phi^2\big) \big(3 \nu (5\nu+2) +
   2 \beta (4\nu^2+13\nu+3) b^2 \tilde\phi^2\big)b'^2\cr
&&-(\nu+1)^2 (3 \nu + 2 \beta (4\nu+1) b^2 \tilde \phi^2)^2b'^4\Big\}
\eea
As a concrete example, we consider $\nu=1$ and $\Lambda_0=0$, and the domain-wall solution is
\be
ds^2=\fft{dz^2 + dx_1^2 + dx_2^2 + dx_3^2 -dt^2}{1+\tilde z^2}\,,\qquad
A=\sqrt{\fft{\tilde z^2(15 + 20 \tilde z^2 + 8\tilde z^4)}{10\beta(1 + \tilde z^2)}}\, dt\,,
\ee
where $\tilde z=\mu\,z$. In this case, the Schr\"odinger potential is
\be
V=\fft{\mu^2(126 + 603 \tilde z^2 + 1368 \tilde z^4 + 1664 \tilde z^6 +
960 \tilde z^8 + 192 \tilde z^{10})}{8 (3 + 15 \tilde z^2 + 20 \tilde z^4 + 8 \tilde z^6)^2}\,.
\ee
Thus we see that the potential is positive definite with a maximum at $\tilde z=0$ and it approaches zero as $\tilde z\rightarrow \pm \infty$. For general $\nu$, there are no analytic expressions for $b(z)$ and $\tilde \phi(z)$; however, since $V$ is a scalar quantity, and we can expressed it in terms of the $r$ coordinate, given by
\bea
V &=& \fft{1}{8\nu^2\big(3\nu + 2\beta (4\nu+1)\phi^2\big)^2a'^2}\Big\{
-4 \mu^4 \nu^4 \beta(4\nu+1)^2 a^4 \phi^4\cr
&& +2 \mu^2 \nu^2 a^2 \big(3 \nu + 2 \beta(4\nu+1) \phi^2\big) \big(3 \nu (5\nu+2) +
   2 \beta (4\nu^2+13\nu +3) \phi^2\big)\cr
 &&- (\nu+1)^2 (3 \nu + 2\beta(4\nu+1) \phi^2)^2 a'^4\Big\}\,.
\eea
It is straightforward to verify that this Schr\"odinger potential for $\nu>0$ will not be able to give rise to a bound state.  As a contrast, the minimally-coupled massless scalar that satisfies the Laplacian equation $\Box_5\Phi=0$ gives rise to the Schr\"odinger equation of the type
\be
V=\ft38 (3a'^2 + 2a a'')=\fft{3\mu^2\big(5\nu(\cosh(2\mu r)-1) - 4\big)}{16 \nu [\cosh(\mu r)]^{2(\nu+1)}}\,.
\ee
which would give a bound state of $p_\mu p^\mu=0$.

Thus although the metric is a smooth realization of the RSII scenario, it does not, at the first sight,  appear to trap gravity on the brane, since the wave function of the Schr\"odinger system is not normalizable. However, this analysis only shows that there is no bound state with $m=0$ in the Schr\"odinger equation (\ref{schr}).  A proper measure whether the $m=0$ graviton state is bounded or not perhaps should be evaluated from the $D=5$ point of view.  The massless graviton mode of the $D=4$ brane world is given by
\be
\xi_{\mu\nu} \Psi_0=\xi_{\mu\nu} e^{{\rm i} p\cdot x}\,,\qquad
\hbox{with}\qquad p^\mu p_\mu=p^\mu \xi_{\mu\nu}=\xi^\mu{}_\mu=0\,.
\ee
Thus from the $D=5$ point of view, we have
\be
\int dr \sqrt{-g}\,|\Psi_0|^2=\int_{-\infty}^\infty dr\, [\cosh(\mu r)]^{-4\nu}\,,
\ee
which is clearly finite for positive $\nu$. It follows that $\Psi_0$ {\it is} normalizable. We may also regard the domain-wall ansatz (\ref{dwans}) as that of the warped Kaluza-Klein reduction, if we replace the Minkowski spacetime $\eta_{\mu\nu} dx^\mu dx^\nu$ with the generic metric
$g_{\mu\nu}(x) dx^\mu dx^\nu$.  The equations of motion in $D=5$ are satisfied provided that the $D=4$ curvature $G_{\mu\nu}^{(4)}=0$, implying the effective Lagrangian in $D=4$ is simply
\be
{\cal L}_4=\sqrt{-g^{(4)}} R^{(4)}\,.\label{d4lag}
\ee
(Here quantities with the superscript ``$(4)$'' depend on the $D=4$ brane coordinates $x^\mu$ only.) The reduction however is not ``consistent'' in that substituting the ansatz into the $D=5$ Lagrangian gives rise to also a cosmological constant in addition to (\ref{d4lag}).  It is nevertheless instructive to plug the Randall-Sundrum domain wall with the generic Ricci-flat world volume into the $D=5$ action and integrate out the $r$ direction.  We find that the result is convergent, given by
\bea
\int_{-\infty}^{\infty} dr\sqrt{-g} L &\sim&  \int_{-\infty}^{\infty}\fft{\mu^2 \nu (5\nu\cosh(2\mu r)-5\nu -4)}{[\cosh(\mu r)]^{2+4\nu}}\cr
&=& \ft{6\mu\nu}{2\nu+1} \big({}_2F_1[1,-2\nu-1;2\nu+1; -1]-1\big)\,.
\eea
It is finite for positive $\nu$.  This is strongly indicative that the four-dimensional graviton is trapped on the brane.

Finally we would like to study whether the tensor perturbation is stable by examining the sign of its kinetic term, which is given by
\be
L\sim \ft12K_0\dot \psi^2 + \cdots\,,\qquad K_0= a^2 \Big(1 -\fft{2\beta (1 +\nu)}{3\nu} \phi^2\Big)\,.
\ee
The ghost-free condition of $K_0\ge 0$ and the reality condition $\phi^2\ge 0$ must be simultaneously satisfied.  This can be indeed achieved.  We consider a concrete representative example, with
\be
\nu=1\,,\qquad \Lambda_0=\ft65\mu^2\,,\qquad \psi_0=0\,.
\ee
In this case $\phi^2=-3/(10\beta)$ which is constant and positive for $\beta\le 0$. In this case $K_0=7a^2/5$, which is also positive definite.  Another potential instability may be caused by the tachyon mode of the Proca field since $\mu_{\rm eff}^2=-148\beta/(45\Lambda_{\rm eff})<0$, which is negative for $\beta<0$.  However, the $\beta<0$ is so far a free parameter and we can easily choose $-15/74\le \beta<0$ so that the Breitenlohner-Freedman bound is satisfied.  For $\nu=1$, both $\phi^2$ and $K_0$ are positive provided that $\Lambda_0>6\mu^2/5$ and $\beta_{\rm min}\le \beta<0$. Analogous conclusions can also be made for other values of $\nu$.

\section{Conclusions}

In this paper, we generalized the Einstein-Maxwell gravity by introducing additional couplings between the vector $A$ with the curvature tensors.  An immediate effect is that the system no longer has the $U(1)$ gauge symmetry associated with the Maxwell field.  However with appropriate couplings, the gauge symmetry can emerge at the linear perturbative level in any background where $A$ vanishes.  Since Maxwell theory is itself a linear theory and furthermore its contribution to the spacetime curvature is negligible, this proposal of the $U(1)$ gauge invariance of the Maxwell theory as an approximate and emergent symmetry can have very little experimental or observational consequence in any scale no more than the solar system.  On the other hand, in the large scale or in cosmology, this proposal can have many non-trivial effects.  We argued that the cosmological constant can provide a global and hence Lorentz invariant mass term to the Maxwell field.  The distribution of dark matter can give rise to a position-dependent (Lorentz violating) term that breaks the gauge symmetry.  Both the global and local symmetry-breaking terms are small, one at the order of the cosmological constant, and the other is proportional to the density of dark matter. In cosmology, the Maxwell field with $U(1)$ gauge symmetry will break the cosmic principle of homogeneity and isotropy. In our Einstein-vector gravity, the vector has no $U(1)$ symmetry and hence it contains a longitudinal scalar mode.  Thus we may use the vector as an inflaton.  We find that inflation is indeed possible and furthermore there exist solutions in which the inflaton vanishes at the late stage.  This allows the emergency of the $U(1)$ gauge symmetry at the linear perturbative level.  Thus the theories are very economical with field contents and address the coming and going of the inflaton.

We constructed a class of Einstein-vector gravities where the vector couples the curvature tensors in any polynomial orders in a bilinear fashion, so that the equation of motion for the vector field is linear in $A$.  We found two specific curvature polynomial combinations, which we call $G^{(k)}$ and $\widetilde G^{(k)}$, such that only Riemann tensor polynomials rather than any of its derivatives appear in the Einstein equations of motion.  This is analogous to the Lovelock gravities and has the effect of remove any ghost excitations associated with higher derivatives acting on a single field.  Thus the theory can be ghost free. In particular in a background of Einstein metric with vanishing $A$, the system may have only one graviton and one photon mode at the linearized level for appropriate choice parameters.

For simplicity, we focused our study on the case where the theory has at most two-derivatives. Some aspects of the cosmological implications of such this theory were studied previously in \cite{Jimenez:2009ai,BeltranJimenez:2010uh}. We obtained many examples of black holes that are asymptotic to Minkowski and Lifshitz spacetimes. We also constructed exact solutions describing wormholes that connect two Minkowski or (A)dS spacetimes.  We constructed domain-wall solutions of the Randall-Sundrum II scenario.  We argued that the domain walls can trap gravity and show that for some appropriate choice of parameters, the kinetic term for the tensor perturbation is positive definite.

We obtain cosmological solutions and we find that the universe can expand with an acceleration without a bared cosmological constant. We also find one class of cosmological solutions that describe de Sitter bounce. As the comoving time $t\rightarrow \pm \infty$, the spacetime becomes the maximally-symmetric de Sitter vacuum with $A=0$.  Furthermore, the effective mass (\ref{mueff}) of the vector field vanishes identically in this case and hence the $U(1)$ gauge symmetry emerges.  The linear spectrum thus contains only the massless graviton and photon modes in later time, as in the Einstein-Maxwell theory. However, at the time when the bounce takes place, the spectrum contains ghost excitations. Furthermore, the new issue of how the inflation ends does not appear to have a resolution within the framework of the model itself.

To conclude, although in our Einstein-vector gravities, the gauge symmetry is violated, it emerges in the background of our current universe, and it can emerge dynamically in the late stage of cosmological evolution.  Furthermore, the existence of de Sitter bounce seems to indicate a possible resolution of the cosmic singularity.  The possible violation owing to cosmological constant and dark matter is sufficiently small so that our proposal cannot be instantly ruled out. Although our theories are rather simple with quadratic $A$, they admits a variety of exact solutions. Furthermore, our theories can be straightforwardly generalized to include higher-order curvature terms. These features make our theories interesting for further investigation.  It is also of interest to investigate the possibility to generalize the vector to include Yang-Mills group indices.  Furthermore, all our discussions, including those on the emerging gauge symmetries, are classically based. It is of interest to investigate the quantum effects.

\section*{Acknowledgement}

We are grateful to Ming-Zhe Li and Yu-Xiao Liu for useful discussions.  The work is supported in part by NSFC grants NO. 11175269, NO. 11475024 and NO. 11235003.
H.L.~would like to thank the participants of the advanced workshop ``Dark Energy and Fundamental Theory'' supported by the Special Fund for Theoretical Physics from the National Natural Science Foundations of China with Grant No. 11447613 for useful discussions.

\appendix
\section{Solutions in general $D$ dimensions}

In the main text, we considered the following general Lagrangian
\be
{\cal L}=\sqrt{-g} \Big( R - 2\Lambda_0 - \ft14 F^2 -\ft12 \mu_0^2 A^2 + \beta A^2 R + \gamma G_{\mu\nu} A^\mu A^\nu -\ft14 {\cal F}^2\Big)\,,\label{mostgenlagAA}
\ee
where $F=dA$, ${\cal F}=d{\cal A}$, $A^2=A^\mu A_\mu$, $F^2=F^{\mu\nu}F_{\mu\nu}$, ${\cal A}^2={\cal A}^\mu {\cal A}_\mu$ and ${\cal F}^2={\cal F}^{\mu\nu}F_{\mu\nu}$.  We have focused on our discussions in four (and five) dimensions.  In this appendix, we present solutions in general $D$ dimensions. The general static ansatz is
\be
ds^2=-a(r)^2 dt^2 + b(r)^2 d\Omega_{n,\epsilon}^2 + c(r)^2 dr^2 \,,
\qquad A=\phi(r)\, dt\,,\qquad {\cal A}=\varphi(r)\, dt\,,
\ee
where $n=D-2$ and $d\Omega_{n,\epsilon}^2$ with $\epsilon=1,0,-1$ is the metric
for the unit $S^{n}$, the $n$-torus or the unit hyperbolic $n$-space.
It is convenient to take $d\Omega_{n,\epsilon}^2=\bar g_{ij} dy^i dy^j$
for general values of $\epsilon$ to be the metric of constant curvature such that its
Ricci tensor is given by $\bar R_{ij}= (n-1)\, \epsilon\, \bar g_{ij}$.  We
may, for example, take $d\Omega_{n,\epsilon}^2$ to be given by
\be
d\Omega_{n,\epsilon}^2 = \fft{du^2}{1-\epsilon u^2} +
    u^2\, d\Omega_{n-1}^2\,,
\ee
where $d\Omega_{n-1}^2$ is the metric of the unit $(n-1)$-sphere.  The basic ingredients of the reduced effective one-dimensional Lagrangian for (\ref{mostgenlagAA}) are
\bea
&&R =  \frac{n (n - 1) \epsilon}{b^2} + \frac{2 a' c'}{a c^3} + \frac{2 n c'
  b'}{c^3 b} - \frac{2 n a' n'}{a c^2 b} - \frac{n (n - 1) b'^2}{c^2 b^2} -
  \frac{2 a''}{a c^2} - \frac{2 n b''}{c^2 b}\,,\cr
&&G_{\mu \nu} A^{\mu} A^{\nu} = \frac{n (n - 1) \epsilon\, \phi^2}{2
  a^2 b^2} + \frac{n \phi^2 c' b'}{a^2 c^3 b} - \frac{n (n - 1) \phi^2 b'^2}{2
  a^2 b^2 c^2} - \frac{n \phi^2 b''}{a^2 c^2 b}\,,\cr
&&A^2 = - \frac{\phi^2}{a^2}\,,\qquad F^2 =-\frac{2 \phi'^2}{a^2c^2},\qquad {\cal F}^2=
-\frac{2 \varphi'^2}{a^2c^2}\,,\qquad \sqrt{- g} = a c b^n .
\eea
The full set of equations of motion associated with $(a,b,c,\phi,\varphi)$ can then be derived from the variation of the reduced Lagrangian.  In particular, $c$ is a pure gauge and can be set to any non-vanishing value or function.  The corresponding equation can be viewed as the vanishing Hamiltonian constraint.

We shall also construct cosmological solutions in general $D=n+1$ dimensions, with the FRLW ansatz
\be
ds^2 =-c(t)^2 dt^2 + a(t)^2 (dx_1^2 + dx_2^2 + \cdots+ dx_n^2)\,,\qquad A=\phi(t) dt\,.\label{gencosmo}
\ee
The ingredients for constructing the explicit reduced Lagrangian are
\bea
&& R  =  - \frac{2 n a' c'}{a c^3} + \frac{n (n - 1) a'^2}{a^2 c^2} + \frac{2
  n a''}{a c^2}\,,\qquad \sqrt{- g} =  c a^n \cr
&&G_{\mu \nu} A^{\mu} A^{\nu}  =  \frac{n (n - 1) a'^2 \phi^2}{2
  a^2 c^4},\qquad
  A^2 =  - \frac{\phi^2}{c^2}\,.
\end{eqnarray}
For the domain-wall ansatz, we can set $c^2=-1$ in (\ref{gencosmo}) and also replace $t$ by $r$ and then $dx_n^2$ by $-dt^2$.

\subsection{Asymptotically-flat black holes}

In this subsection, we generalize the exact special solution of the asymptotically-flat black hole of section \ref{bhexact}.  The existence of the solution requires that $\Lambda_0=\mu^2=\beta=0$, ${\cal A}=0$  and
\be
\gamma=\fft{n-1}{2n}\,.
\ee
In other words, it is the theory (\ref{lag1}) with the above $\gamma$ constraint. The solution is given by
\be
ds^2 = -f dt^2 + \fft{dr^2}{f} + r^2 d\Omega_n^2\,,\qquad  A=\phi\, dt\,,
\ee
\be
f=1 - \big(\fft{r_0}{r}\big)^{\fft12(n-1)}\,,\qquad \phi = \sqrt{\ft{4n}{n-1}}\, f\,.
\ee

\subsection{Wormholes}

In this subsection we generalize the wormhole solutions in section \ref{whsol} to general $D=n+2$ dimensions.  This requires setting $\beta=0$, $\mu_0^2=-2\gamma \Lambda_0$ and ${\cal A}=0$.  In other words, it is the theory (\ref{lag2}).  We make a gauge choice
\be
c=ab^n\,.
\ee
The effective one-dimensional Lagrangian is given by $L=T-V$ with
\bea
T &=& \big(2-\fft{\gamma \phi^2}{a^2}\big)\fft{na' b'}{ab} + \big(1 + \fft{\gamma\phi^2}{2a^2}\big)
\fft{n(n-1)b'^2}{b^2} + \fft{n \gamma (\phi^2)' b'}{a^2b} + \fft{\phi'^2}{2a^2}\,,\cr
V &=&\ft12 b^{2n-2} (-2\epsilon n(n-1) a^2 + 4 \Lambda_0 a^2 b^2 + 2\gamma \Lambda_0 b^2 \phi^2 -\gamma \epsilon n(n-1)\phi^2)\,.
\eea
Following the procedure spelled out in section \ref{whsol}, we find that there exists a superpotential
\be
W=\fft{n}{a} (2a^2 + \gamma\phi^2) \sqrt{b^{n-1} (-m + \epsilon\, b^{n-1} -
\ft{2\Lambda_0}{n(n+1)} b^{n+1})}\,,
\ee
where $m$ is an arbitrary constant. This implies that the full second-order equations can be solved by the following first order equations
\bea
a' &=& \fft{\big(n(n-1)(n+1)m - 4\Lambda_0 b^{n+1}\big) (2a^2 + \gamma \phi^2)^2}{
2(n+1) W b(2a^2 - \gamma\phi^2)}\,,\cr
b' &=& -\fft{\big(n(n+1)(m-\epsilon b^{n-1}) + 2\Lambda_0 b^{n+1}\big) (2a^2 + \gamma\phi^2)}{ (n+1) aW}\,,\cr
\phi' &=& 0\,.
\eea
These equations can be solved straightforwardly and we have
\bea
ds^2 &=& - h dt^2 + \fft{dr^2}{f} + r^2 d\Omega_{n,\epsilon}^2\,,\qquad A=\sqrt{-\ft{2h_0}{\gamma}}\, dt\,,\cr
f &=& -\ft{2\Lambda_0}{n(n+1)} r^2 + \epsilon - \fft{m}{r^{n-1}}\,,\qquad
h=h_0 + \ft12 \Big(1 + \sqrt{1 + \fft{4h_0}{f}}\Big)f\,.
\eea
The reality condition requires that $\gamma<0$ and the solutions describe asymptotic (A)dS wormholes.

When $\Lambda_0=0$, we require that $\epsilon=1$ and make a convention that $h\sim f=1$ as $r\rightarrow \infty$, we find that the $f$ is given by
\be
f=1 - \fft{m}{r^{n-1}}\,,
\ee
with $h$ given by (\ref{hworm}).

\subsection{Lifshitz black holes}

In this subsection of the appendix, we generalize the Lifshitz black holes constructed in sections \ref{lbh} and \ref{clbh} from four dimensions to general $D=n+2$ dimensions.  We shall skip the details of the construction and simply present the solutions.  The exact solutions for neutral Lifshitz black holes exist for all $z$, given by
\bea
ds^2 &=& \ell^2 \Big(-r^{2z}\, f\, dt^2 + \fft{dr^2}{r^2f} + r^2 (dx_1^2 + \cdots + dx_n^2)\Big)\,,\qquad A=q r^z f\, \ell dt\,,\cr
f&=& 1 - \big(\fft{r_0}{r}\big)^{\fft12(z+n)}\,,\qquad
q^2=\fft{4(z-1)(n+2-z)(n+3z)}{z(n+z)(n+8+z)}\,.
\eea
The parameters of the theories are required to be
\bea
\Lambda_0 &=& \fft{(n+z)\big((n+4)(n-1) + 2z(n+3-z)\big)}{2(n+8+z)\ell^2}\,,\qquad
\beta=-\fft{n+z^2-z}{2(z-1)(n+3z)}\,,\cr
\mu_0^2 &=& \fft{z(n+z)\big(n^3+n(3z-1) - n z (2z-3) - 4 z(z^2-3z+2)\big)}{
2\ell^2(z-1)(n+2-z)(n+3z)}\,.
\eea
Note that in these solution the parameter $\beta$ cannot be zero.

Exact solutions of charged Lifshitz black holes under the Maxwell field ${\cal A}$ exist only for $z=3n$, and they are given by
\bea
ds^2 &=& \ell^2 \Big(-r^{6n}\, f\, dt^2 + \fft{dr^2}{r^2f} + r^2 (dx_1^2 + \cdots + dx_n^2)\Big)\,,\cr
A &=& \sqrt{\ft{(3n-1)(n-1)}{n(3n-2+8\beta(3n-1))}}\,r^{3n}\, f\, \ell dt\,,\qquad
{\cal A} = (\varphi_0 + Q r^{2n})\, \ell dt\,,\cr
f&=& 1 - \fft{\big(16n(3n-1)\beta +2n(3n-2)\big) Q^2}{(n-1)\big(20(3n-1)\beta + 9n-2\big)
r^{2n}}\,,\cr
\Lambda_0 &=& \fft{(n-1)\big(16(3n-1)(2n-1)\beta + 21n^2-23n+4\big)\mu_0^2}{
512n(3n-1)(2n-1)\beta^2 + 32n(9n^2-7n+2)\beta - 4n (3n-2)(n-2)}\,,\cr
\ell^2 &=& \fft{2n^2\big(n-2 - 16\beta (2n-1)\big)}{(n-1)\mu_0^2}\,,\qquad
\gamma=-\fft{2(7n+1)\beta+n}{n-1}\,.
\eea
In these solutions, the parameter $\beta$ can be set to zero.

\subsection{de Sitter bounce}

Here we generalize the four-dimensional de Sitter bounce solutions in section \ref{cosmo} to general $D=n+1$ dimensions:
\bea
ds_{n+1}^2&=& -dt^2 + [\cosh(\mu t)]^\nu (dx_1^2 + dx_2^2 + \cdots + dx_n^2)\,,\qquad
A=\phi\, dt\,,\cr
\phi^2 &=& \ft{(n-1)\nu}{2\beta}[\cosh(\mu t)]^{1-n\nu}\, \psi\,,\cr
\psi &=& \sinh(\mu t)\,\psi_0 + \sinh^2(\mu t)\, {}_2F_1[\ft12,
\ft12(3-n\nu); \ft32; -\sinh^2(\mu t)]\cr
 &&+ \fft{\Lambda_0}{\Lambda_{\rm eff}} \,{}_2F_1[-\ft12,\ft12(1-n\nu); \ft12; -\sinh^2(\mu t)]\,,
\eea
where $\Lambda_{\rm eff}=\fft12 n(n-1)\mu^2\nu^2$. The parameters $(\mu,\nu)$ are related to those in the original theory as
\be
\gamma=\fft{2\beta\big( (n+1)\nu-2\big)}{(n-1)\nu}\,,\qquad
\mu_0^2=4n\beta\mu^2\nu\,.
\ee

\subsection{Domain walls}

Here we generalize the five-dimensional domain walls in section \ref{sec:dw} to general $D=n+1$ dimensions :
\bea
ds_{n+1}^2&=& dr^2 + \fft{1}{[\cosh(\mu r)]^\nu} (-dt^2 + dx_1^2 + \cdots + dx_{n-1}^2)\,,\qquad
A=\phi\, dr\,,\cr
\phi^2 &=& \ft{(n-1)\nu}{2\beta} [\cosh(\mu r)]^{1+n\nu} \psi\,,\cr
\psi &=& \sinh(\mu r)\,\psi_0 + \sinh^2(\mu r)\, {}_2F_1[\ft12,
\ft12(3+n\nu); \ft32; -\sinh^2(\mu r)]\cr
 &&+ \fft{\Lambda_0}{\Lambda_{\rm eff}} \,{}_2F_1[-\ft12,\ft12(1+n\nu); \ft12; -\sinh^2(\mu r)]\,,
\eea
where $\Lambda_{\rm eff}=-\fft12n(n-1)\mu^2\nu^2$. The parameters $(\mu,\nu)$ are related to those in the original theory as
\be
\gamma=\fft{2\beta\big( (n+1)\nu+2\big)}{(n-1)\nu}\,,\qquad
\mu_0^2=4n\beta\mu^2\nu\,.
\ee

\end{document}